\documentclass[aps,reprint,superscriptaddress,nofootinbib,longbibliography]{revtex4-2}
\usepackage{graphicx, color, soul}
\usepackage{xcolor, bbold}      
\usepackage{siunitx}
\usepackage{amsmath, amssymb}
\usepackage{braket}
\usepackage{color,soul,url}
\usepackage{float}
\usepackage[hidelinks]{hyperref}
\usepackage{microtype} 
\usepackage[]{newtxtext} 
\usepackage[]{newtxmath}

\usepackage{hyperref}
\hypersetup{colorlinks,
	linkcolor={blue!75!black!80!yellow},
	citecolor={blue!75!black!80!yellow},
	urlcolor={blue!75!black!80!yellow}
}

\usepackage{booktabs}
\usepackage{colortbl}
\newcommand{\separatorrule}{\arrayrulecolor{black!25}\specialrule{.25pt}{.65\jot}{.65\jot}\arrayrulecolor{black}}
\usepackage{makecell}

\newcommand{\vk}{\Pi}
\newcommand{\vkk}{{\bf \Pi}}
\newcommand{\ii}{\mathrm{i}}
\newcommand{\tr}{\text{Tr}} 
\newcommand{\A}{\mathcal{A}} 

\newcommand{\matrixel}[3]{\left<#1 \vphantom{#2} \vphantom{#3} \left| #2 \vphantom{#1} \vphantom{#3} \right| #3 \vphantom{#1} \vphantom{#2} \right>}

\newcommand{\raisedchi}{\raisebox{\depth}{\(\chi\)}}

\begin{document}
\title{Response to polarization and weak topology in Chern insulators}

\author{Sachin Vaidya}
\email{sachin4594@gmail.com}
\affiliation{Department of Physics, The Pennsylvania State University, University Park, PA 16802, USA}
\affiliation{Department of Physics, Massachusetts Institute of Technology, Cambridge, Massachusetts 02139, USA}

\author{Mikael C. Rechtsman}
\affiliation{Department of Physics, The Pennsylvania State University, University Park, PA 16802, USA}
\author{Wladimir A. Benalcazar}
\email{benalcazar@emory.edu}
\affiliation{Department of Physics, Emory University, Atlanta, GA 30322, USA}

\date{\today}

\begin{abstract}
Chern insulators present a topological obstruction to a smooth gauge in their Bloch wave functions that prevents the construction of exponentially-localized Wannier functions - this makes the electric polarization ill-defined. Here, we show that spatial or temporal differences in polarization within Chern insulators are well-defined and physically meaningful because they account for bound charges and adiabatic currents. We further show that the difference in polarization across Chern-insulator regions can be quantized in the presence of crystalline symmetries, leading to ``weak'' symmetry-protected topological phases. These phases exhibit charge fractional quantization at the edge and corner interfaces and with concomitant topological states. We also generalize our findings to quantum spin-Hall insulators and 3D topological insulators. Our work settles a long-standing question and deems the bulk polarization as the fundamental quantity with a ``bulk-boundary correspondence'', regardless of whether a Wannier representation is possible.
\end{abstract}
\maketitle

The concept of electric polarization is essential in describing insulating materials and is at the core of our understanding of topological phases of matter. Although heuristically understood as the dipole moment per unit volume, its determination in crystalline materials is subtle \cite{Vanderbiltbook, resta2007theory}. In the 1990s, a correct definition of polarization was formulated in terms of the (gauge-invariant) Berry phase of the Bloch wave functions across the Brillouin zone~\cite{zak1989, resta1994macroscopic, Vanderbiltbook, resta2007theory, vanderbilt1993electric, king1993theory}. The Berry phase encodes the positions of the spatially-resolved Wannier functions, so-called ``Wannier centers'', which further facilitates establishing the bulk-boundary correspondence for polarization in crystals. However, the recent explorations of topological insulators has made it apparent that a Wannier representation is not always possible -- such is the case of Chern insulators~\cite{qi2006topological, haldane1988model} -- and thus, whether the concept of polarization can be extended beyond the Wannier center picture has gained relevance and remains an open question in topological band theory.

In this work, we show that the concept of polarization in Chern insulators, although mathematically ill-defined in the bulk, is physically meaningful. This is because the physical manifestations of polarization, namely electronic bound charges and adiabatic currents, are proportional to \emph{changes in polarization} (not the polarization itself), and these changes can be well-defined in Chern insulators. Additionally, we show that under crystalline symmetries, the bound charge is fractionally quantized, as in conventional insulators. This leads to the emergence of weak symmetry-protected topological phases within each Chern class, which can host edge states, and also corner states akin to those found in higher-order topological phases. We present our results in a tight-binding model and an experimentally-realizable microwave photonic crystal. 

In conventional insulators, the electric polarization ${\bf P}$ is defined in terms of Berry phases along the non-contractible loops of the Brillouin zone \cite{resta1994macroscopic} or equivalently in terms of the Wannier centers of the occupied bands \cite{Vanderbiltbook, marzari2012maximally, brouder2007exponential}. In 2D crystalline insulators, the bulk polarization ${\bf P}=P_1 {\bf a}_1+P_2 {\bf a}_2$, where ${\bf a}_{i=1,2}$ are primitive lattice vectors, has components $P_i = \oint \mathrm{d}^2{\bf k} \tr[\A_i({\bf k})]$, where $\A_i$ is the Berry connection, with elements $[\A_i({\bf k})]_{m,n}=-\ii \matrixel{u_m({\bf k})}{\partial_{k_i}}{u_n({\bf k})}$, and $\ket{u_m({\bf k})}$ is the Bloch eigenstate of occupied band $m$ at crystal momentum ${\bf k}=(k_1,k_2)$. The components of $\mathbf{P}$ can also be written as $P_i =\oint \mathrm{d}k_j p_i(k_j)$, where
\begin{align}
p_i(k_j) &= \frac{1}{2\pi} \oint \mathrm{d}k_i \tr[\A_i({\bf k})] \;\; \mbox{ mod 1} \label{eq:P}
\end{align}
is the $k_j$-sector polarization, for $i,j=1,2$; $i \neq j$.

Chern insulators are paradigmatic topological materials. They are insulating in the bulk but have conducting chiral edge states~\cite{qi2006topological, haldane1988model}. In Chern insulators, $p_i(k_j)$ winds around the 1D Brillouin zone formed by $k_j \in [-\pi/a_j,\pi/a_j)$. This winding simultaneously reflects the difficulty in building exponentially-localized Wannier functions \cite{thonhauser2006insulator, thouless1984wannier} and defining the bulk polarization, as the value of $P_i$ depends on the starting point in the loop integral along $k_j$~\cite{VanderbiltPolarizationChern2009}. Furthermore, the chiral edge states that cross the Fermi level also complicate establishing the bulk-boundary correspondence to polarization because the bound charge, if it exists, would be affected by the partial occupation of its chiral edge states \cite{VanderbiltPolarizationChern2009}. 

Our starting point to address the question of polarization in Chern insulators is to focus on its associated physical observables. Consider the interface between two regions, $R_1$ and $R_2$. A charge density $\sigma$ arises due to the \emph{difference in polarization} across this interface, following the ``interface-charge theorem''~\cite{Vanderbiltbook, vanderbilt1993electric},
\begin{align}
    \sigma = \left[{\bf P}^{(R_1)} - {\bf P}^{(R_2)} \right] \cdot {\bf \hat{n}} \quad \text{mod }1,
    \label{eq:interface_charge_theorem}
\end{align}
where we have set the unit cell lengths in all directions and the electronic charge to unity for simplicity, and ${\bf \hat{n}}$ is the vector normal to the surface. If regions $R_1$ and $R_2$ have inequal Chern numbers, i.e., $C_1 \neq C_2$, $|C_1-C_2|$ chiral edge states will appear at their common interface, rendering it metallic. This, in conjunction with the aforementioned winding of $p_i(k_j)$, makes the definition of polarization problematic. Coh and Vanderbilt studied how a definition of the polarization might be saved in the case $C_1=1$, $C_2=0$, but only with the knowledge of the wave vector at which the (partial) occupancy of the edge state is discontinuous \cite{VanderbiltPolarizationChern2009}.

Here, we instead consider the case in which $C_1=C_2$, so that $p^{(\alpha)}_i(k_j)$ (Eq.~\ref{eq:P}), for regions $\alpha=\{ R_1, R_2 \}$, individually wind, but where the difference in $k$-sector polarizations across the two regions
\begin{align}
    \Delta p_i(k_j) = p^{(R_1)}_i(k_j)-p^{(R_2)}_i(k_j)
    \label{eq:DeltaP}
\end{align}
does not wind, and is \emph{non-zero}. The key insight here is that this configuration preserves the non-trivial nature of Chern insulators but allows for insulating interfaces within them that can be probed for responses to spatial changes in polarization. As we shall see, this insight will enable a physical notion of polarization that yields measurable observables, i.e. bound charges consistent with Eq.~\eqref{eq:interface_charge_theorem}. We will also demonstrate that adiabatic variations of the polarization within Chern insulators can result in pumping of charges via adiabatic currents. Taken together, both observables (bound charges and currents) are sufficient to demonstrate that crystals without a Wannier representation can exhibit a response to electric polarization. 

\begin{figure}[t]
    \centering
    \includegraphics[width=\columnwidth]{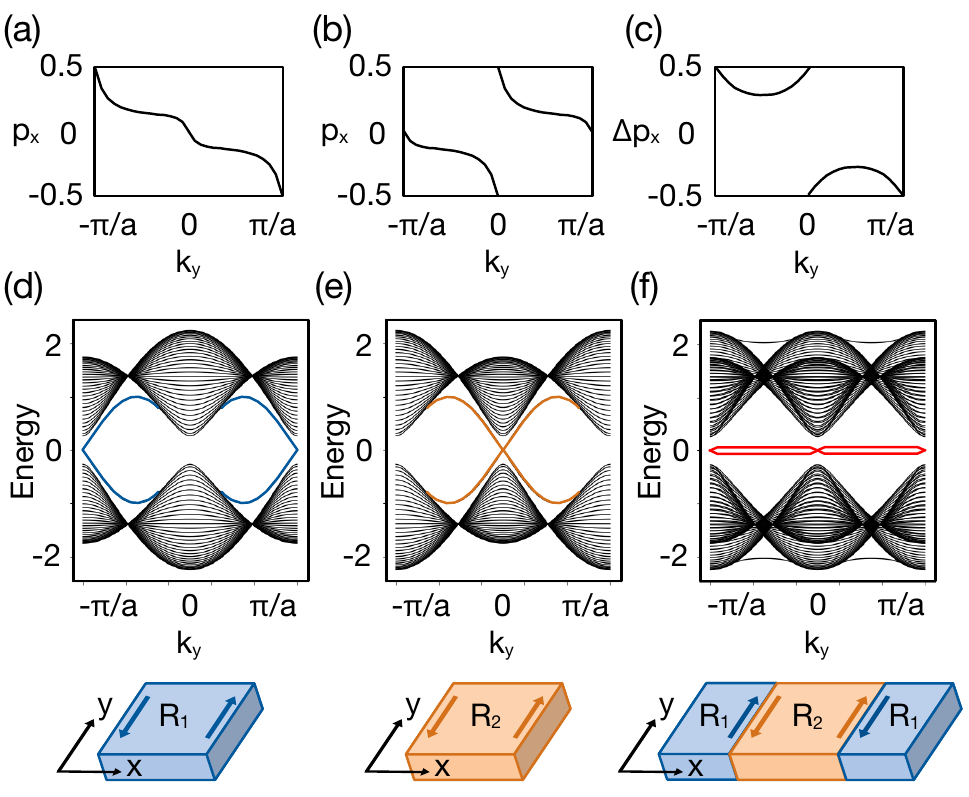}
    \caption{$k$-sector polarization and boundary states in Chern insulators. (a), (b) $p_x(k_y)$ for the occupied band of $h_{\text{QWZ}}(\mathbf{k},0)$ and $h_{\text{QWZ}}(\mathbf{k},\pi)$, respectively, for $m = 0.25$. (c) $\Delta p_x(k_y)$ for $h_{\text{QWZ}}(\mathbf{k},0)$ and $h_{\text{QWZ}}(\mathbf{k},\pi)$. (d), (e) Energy bands for $h_{\text{QWZ}}(\mathbf{k},0)$ and $h_{\text{QWZ}}(\mathbf{k},\pi)$ respectively, with open boundaries as depicted in the bottom panels. The chiral edge states are highlighted in blue and orange. (f) Energy bands for an inversion-symmetric configuration consisting of two regions described by $h_{\text{QWZ}}(\mathbf{k},0)$ and $h_{\text{QWZ}}(\mathbf{k},\pi)$ as depicted in the bottom panel. The system has periodic boundaries imposed along both directions and has two internal interfaces. The non-chiral edge states at these interfaces are highlighted in red.} 
    \label{fig:fig1}
\end{figure}

We first present the accumulation of electronic bound charge in Chern insulators using a two-band tight-binding model described by the following generalized Qi-Wu-Zhang (QWZ) Bloch Hamiltonian,
\begin{align}
h_\mathrm{QWZ}(\mathbf{k},\theta) = \sin k_x \sigma_x + \sin (k_y+\theta) \sigma_y + \nonumber\\ [m + \cos k_x + \cos (k_y+\theta)] \sigma_z,
\label{eq:QWZ}
\end{align}
where $\mathbf{k} = (k_x, k_y)$ is the crystal momentum, $\sigma_{x,y,z}$ are the Pauli matrices and $m$ is a mass term. For $\theta = \theta^* = 0$ and $\pi$, this Hamiltonian possesses inversion symmetry, $\mathcal{I}h({\bf k},\theta^*)\mathcal{I}^{-1}=h(-{\bf k},\theta^*)$, with $\mathcal{I} = \sigma_z$, as well as particle-hole symmetry $\Xi h({\bf k},\theta^*)\Xi^{-1}=-h(-{\bf k},\theta^*)$ with $\Xi = i\sigma_y\mathcal{K}$, where $\mathcal{K}$ is complex-conjugation. The Hamiltonian in \eqref{eq:QWZ} is gapped for all values of ${\bf k}$ and $\theta$. The value of $m$ sets the Chern number, $C$, of the two bands of this model; for the lowest band, $C = 1$ for $0<m<2$, $C = -1$ for $-2<m<0$, and $C = 0$ otherwise. The plots of $p_x(k_y)$ for $m=0.25$ and for $\theta^* = 0$ and $\pi$ exhibit a non-trivial winding due to the non-zero value of $C$ and are shown in Fig.~\ref{fig:fig1}(a) and (b), respectively. Under open boundary conditions along one direction, i.e., with vacuum on the exterior, these systems host chiral edge states as seen from the energy bands in Fig.~\ref{fig:fig1}(d) and (e).

We now consider two adjacent regions, $R_1$ and $R_2$, with Bloch Hamiltonians $h_{R_1}(\mathbf{k})=h_\mathrm{QWZ}(\mathbf{k},0)$ and $h_{R_2}(\mathbf{k})=h_\mathrm{QWZ}(\mathbf{k},\pi)$, respectively. Although $h_{R_1}$ and $h_{R_2}$ have a winding in $p_i(k_j)$, as shown in Fig.~\ref{fig:fig1}(a) and (b), the quantity $p^{(1)}_i(k_j)-p^{(2)}_i(k_j)$ does not wind, as shown in Fig.~\ref{fig:fig1}(c)\footnote{It is assumed here that the two bulk materials share the same periodicity and that the same choice of gauge is made for their Berry connections, $\mathbf{\mathcal{A}}_1$ and $\mathbf{\mathcal{A}}_2$, in {E}q.~\eqref{eq:P}.}. Crucially, the difference in polarization across the interface separating the two regions, $\Delta P_i$, given by 
\begin{align}
\Delta P_i = \oint \mathrm{d}k_j \Delta p_i(k_j),
\end{align} 
is equal to $\frac{1}{2}$.

\begin{figure}[t]
    \centering
    \includegraphics[width=\columnwidth]{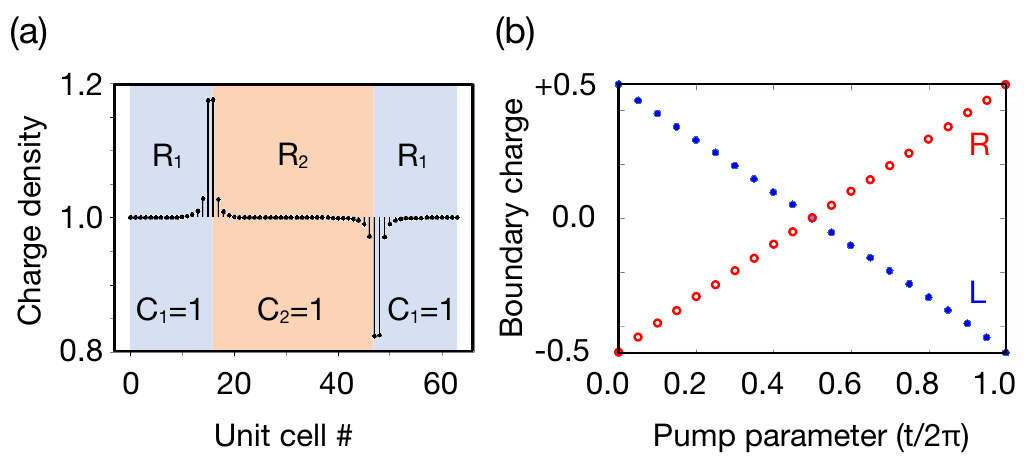}
    \caption{Bound charge and adiabatic current in Chern insulators. (a) The charge density at half filling in the system shown in Fig.~\ref{fig:fig1}(f). Under inversion symmetry, the bound charges that appear at the interfaces are quantized to $\pm 0.5$. (b) The left (L) and right (R) bound charges under an adiabatic change in the polarization as a function of the parameter $t$. The system has inversion symmetry at $t/2\pi = 0$ and $0.5$.} 
    \label{fig:fig2}
\end{figure}

To explore the consequences of this non-zero difference in polarization, we consider a finite, inversion-symmetric system that consists of the two regions described by the Hamiltonians $h_{R_1}$ and $h_{R_2}$ as shown at the bottom of Fig.~\ref{fig:fig1}(f). In this system, we observe the appearance of non-chiral edge states in the energy bands shown in Fig.~\ref{fig:fig1}(f). As a result of this and the presence of inversion symmetry, fractional charge densities (per unit length) quantized to $\pm \frac{1}{2}$ appear at each of the two interfaces at exactly half filling\footnote{An infinitesimal breaking of inversion symmetry is necessary to lift the degeneracy of edge states, only one of which is occupied at half filling. This breaking of inversion symmetry fixes the sign of the fractional charges at each boundary.}, as shown in Fig.~\ref{fig:fig2}(a). This is consistent with the interface charge theorem in Eq.~\eqref{eq:interface_charge_theorem} and is analogous to the expected response to polarization in conventional insulators.

We next turn our attention to the second physical observable associated with polarization - a current density in the bulk that appears due to an adiabatic change in polarization in time. The system considered above for determining the accumulation of bound charge is also useful for probing such adiabatic currents. Since a Wannier center picture is impossible for Chern insulators, the bulk currents are difficult to examine directly. It is possible to visualize these currents using the adiabatic evolution of hybrid Wannier centers~\cite{VanderbiltPolarizationChern2009}; however, this does not yield a clear physical observable. Instead, we probe the existence of these bulk currents via the bound charges that appear at Chern-insulator interfaces. To show this, we adiabatically evolve the region $R_2$ in Fig.~\ref{fig:fig1}(f), by changing the parameter $\theta$ as $\theta(t) = t$, i.e. $h_{R_2}=h_\mathrm{QWZ}(\mathbf{k},t)$ for $t \in [0,2\pi)$, while keeping the region $R_1$ constant, i.e., $h_{R_1}=h_\mathrm{QWZ}(\mathbf{k},0)$. The two regions remain gapped in the bulk for the full cycle of the adiabatic parameter. In Fig.~\ref{fig:fig2}(b), we plot the bound charges as a function of $t$, where we observe the change in bound charge by $\pm 1$ unit of charge. The fact that the pumping of a single charge unit during the cycle is observed at the boundaries, implies by continuity, that this charge was also pumped from the left to the right of each unit cell, giving rise to a current density in the bulk of the Chern insulator region $R_2$. We emphasize that while this last statement would be trivial in the case of a Wannierizable system, it is far from trivial in the absence of a Wannier representation.

\begin{figure*}[ht]
    \centering
    \includegraphics[width=1.9\columnwidth]{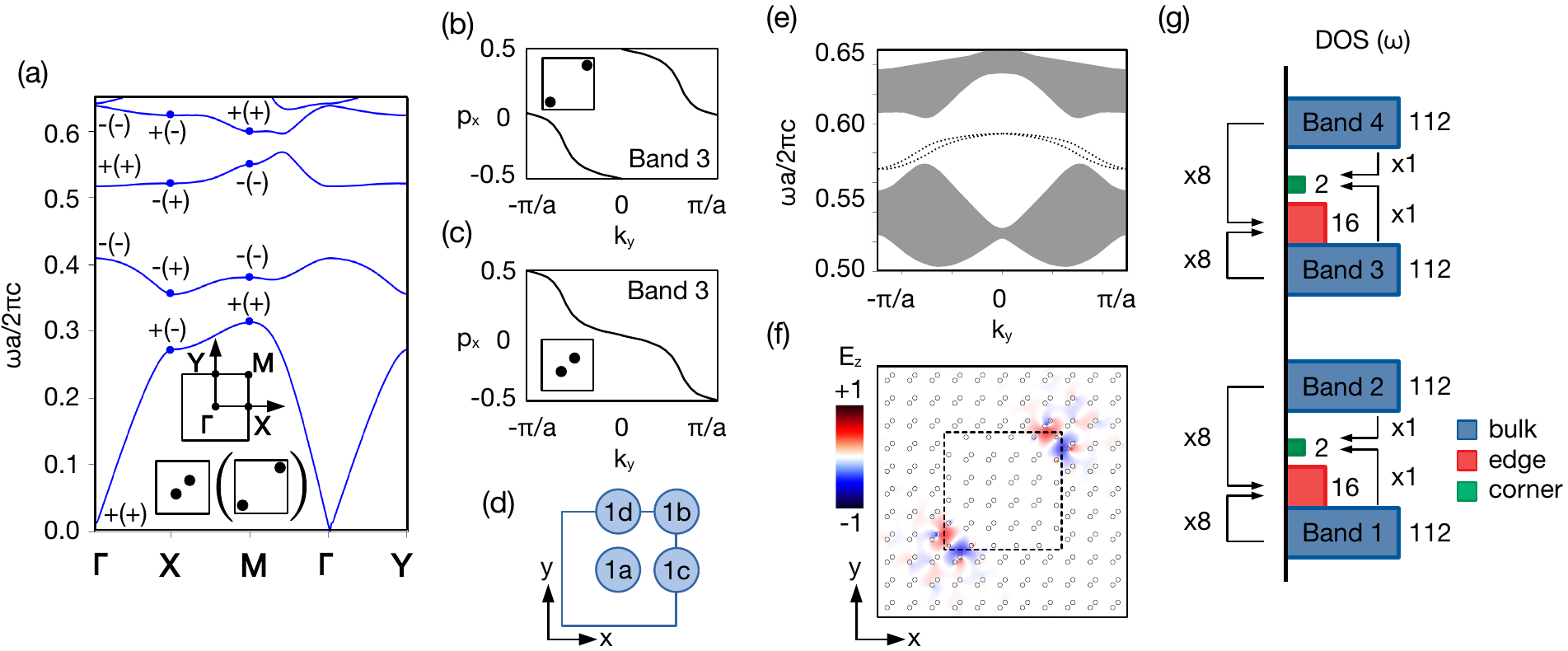}
    \caption{(a) The TM band structure of the PhCs with unit cells shown in the inset. The high-symmetry points are labeled by $\mathcal{I}$-eigenvalues (signs within parenthesis correspond to configuration with unit cell in parenthesis). (b), (c) The Wilson loop spectrum for band 3 for the contracted and expanded unit cells, respectively. (d) Maximal Wyckoff positions in an $\mathcal{I}$-symmetric unit cell. (e) The frequency spectrum of a composite $\mathcal{I}$-symmetric system made of the two types of unit cells with two interfaces as shown in the lower panel of Fig.~\ref{fig:fig1}(f). In-gap edge states (dotted lines) are localized at the interface between the two regions. (f) The $E_z$ mode profile for corner states that arise at the corners between the two Chern PhCs in a ``core-cladding'' geometry. The inner core region (dashed box) consists of the expanded unit-cell type and the outer cladding region consists of the expanded unit-cell type. (g) A schematic of the photonic density of states for TM bands 1 to 4 for the finite system with $11\times 11$ unit cells shown in (f).}
    \label{fig:fig3}
\end{figure*}

We have seen that inversion symmetry quantizes the difference in polarization in Chern insulators to $0$ or $\frac{1}{2}$, and correspondingly the bound charges to $0$ or $\pm \frac{1}{2}$. This implies the existence of well-defined \emph{weak} topological phenomena\footnote{``Weak'' topological phenomena here refers to those protected and enforced by crystalline symmetries (as opposed to the strong  topological phenomena of the tenfold classification).} at the interface between a pair of inversion-symmetric Chern insulators with the same Chern number. To explore this further, we turn to a description of inversion-symmetric insulators using a symmetry-indicator approach.

Crystalline energy bands in class A of the ten-fold classification under inversion symmetry are characterized by a set of invariants given by~\cite{Rot_HOTI1, vaidya2023topological, Wilson_loop_inversionsym, Inv_sym_topo_insulators}
\begin{align}
    \raisedchi = \left( C\, \Big| \, [X], [Y], [M] \right),
\end{align}
where $C$ is the Chern number and the symmetry indicators $[\vk]$ are defined as $[\vk] \equiv \# \vkk-\#\mathbf{\Gamma}$, where $\# \vkk$ is the number of states below the Fermi level with inversion eigenvalue $+1$, at the high-symmetry point (HSP) $\vkk$. The set of single isolated Wannierizable bands can be enumerated exhaustively using the induction of band representations, starting from symmetric Wannier functions at maximal Wyckoff positions~\cite{TQC1, canoEBRs, watanabe2017}. (Table~\ref{tab:C2_Invariants_table}).

Insulators whose occupied bands are characterized by $\raisedchi$ invariants with a vanishing Chern number are either \emph{atomic limits} or \emph{fragile} phases. For such insulators, the polarization (with respect to vacuum) can be calculated as ${\bf P}(\raisedchi)=(1/2)\left( [Y]+[M] \right) {\bf a}_1+(1/2)\left( [X]+[M] \right) {\bf a}_2$, where ${\bf a}_1$ and ${\bf a}_2$ are the primitive lattice vectors~\cite{Rot_HOTI1,vaidya2023topological}. Additionally, some indices can lead to higher-order topological corner states for both atomic limits and fragile phases, determined by the corner charge index $Q(\raisedchi)=(1/4)\left( -[X]-[Y]+[M] \right)$ \cite{Rot_HOTI1,vaidya2023topological}\footnote{Defining $\mathbf{P}$ and $Q$ for fragile phases is a little more subtle. Fragile phases ($F$) have non-Wannierizable bands that can be formally expressed as differences of atomic limits (say $A_1$ and $A_2$) as $F=A_1 \ominus A_2$. Adding the correct atomic-limit degrees of freedom (in this case, $A_2$) to $F$ renders the full set of bands Wannierizable. $\mathbf{P}$ and $Q$ can be calculated for $F$ by first calculating them for the Wannierizable set, $F\oplus A_2 = A_1$, and then removing the contribution from $A_2$.}. Both $\mathbf{P}(\raisedchi)$ and $Q(\raisedchi)$ are defined modulo a unit electronic charge. Table~\ref{tab:C2_Invariants_table} also shows the values of these quantities for the full set of single-band atomic limits under inversion symmetry.

For a pair of Chern-insulator regions described by indices $\raisedchi_1$, $\raisedchi_2$ but with the same Chern number $C_1 = C_2 = C$, the relative index, $\Delta \raisedchi = \raisedchi_2 - \raisedchi_1$, defined as
\begin{align}
\Delta \raisedchi = \left( C_2-C_1\, \Big| \, [X]_2-[X]_1, [Y]_2-[Y]_1, [M]_2-[M]_1 \right),  
\end{align}
has a vanishing Chern component and describes either an atomic limit or a fragile phase. As a result, while $\mathbf{P}(\raisedchi_{\alpha})$ and $Q(\raisedchi_{\alpha})$ for $\alpha \in \{1,2\}$, are ill-defined for the Chern-insulators individually, the difference in polarization, $\Delta \mathbf{P}:= \mathbf{P}(\Delta\raisedchi)$, and relative corner charge, $\Delta Q:= Q(\Delta\raisedchi)$, can be defined. These quantities are associated with physical observables at the interface between the two regions, i.e., edge and corner states appear at the interface between the Chern insulators described by $\raisedchi_1$ and $\raisedchi_2$, with non-zero $\Delta\mathbf{P}$ and $\Delta Q$.

\begin{table}[!htb]
	\centering
		\centering
	\begin{tabular}{ccccc}
		\toprule 
		Wyckoff Pos. & Site symm. & $\raisedchi = \left( C\, \Big| \, [X], [Y], [M] \right)$ & $\mathbf{P}$ & Q\\
		\midrule
		$1a$ & $\rho(\mathcal{I})=\pm 1$ &  $(0\, | \,0,0,0)$ & $\mathbf{0}$ & 0\\
		\separatorrule
        \makecell{$1c$ \\ $1c$} & \makecell{$\rho(\mathcal{I})=+1$ \\ $\rho(\mathcal{I})=-1$} & \makecell{$(0\, | \,-1,0,-1)$ \\ $(0\, | \,1,0,1)$} & $\frac{1}{2} \mathbf{a}_1$ & 0 \\
		\separatorrule
        \makecell{$1d$ \\ $1d$} & \makecell{$\rho(\mathcal{I})=+1$ \\ $\rho(\mathcal{I})=-1$} & \makecell{$(0\, | \,0,-1,-1)$ \\ $(0\, | \,0,1,1)$} & $\frac{1}{2} \mathbf{a}_2$ & 0 \\
		\separatorrule
        \makecell{$1b$ \\ $1b$} & \makecell{$\rho(\mathcal{I})=+1$ \\ $\rho(\mathcal{I})=-1$} & \makecell{$(0\, | \,-1,-1,0)$ \\ $(0\, | \,1,1,0)$} & $\frac{1}{2} \left( \mathbf{a}_1 + \mathbf{a}_2 \right)$ & $\frac{1}{2}$ \\
		\bottomrule
	\end{tabular} 
	\caption{Indices induced from inversion-symmetric Wannier functions centered at all possible maximal Wyckoff positions, shown in the unit cell of Fig.~\ref{fig:fig3}(d). $\rho(\mathcal{I})$ is the inversion symmetry representation of the Wannier function.}
	\label{tab:C2_Invariants_table}
 \end{table}

In the tight-binding example of Fig.~\ref{fig:fig1}(f), we have already seen the appearance of such edge states induced by a non-zero difference in polarization across the boundary. We now show the generality of this framework by demonstrating the presence of these edge states and corner states in experimentally-realizable microwave photonic crystals (PhCs) that do not admit a tight-binding description \cite{photoniccrystalsbook, raghu2008analogs, haldane2008possible, wang2008reflection, wang2009observation}.

The unit cells of the proposed inversion-symmetric, two-dimensional PhCs, with lattice constant $a$, are shown in the inset of Fig.~\ref{fig:fig3}(a), each of which consists of two dielectric discs made out of Yttrium-Iron-Garnet (YIG) ($\varepsilon = 15$), a strong magneto-optical material at microwave frequencies. These two unit cells are related by a $a/2$ shift in both $x$ and $y$ directions, and we refer to them as ``contracted'' and ``expanded''. Time-reversal symmetry is broken by applying a magnetic field in the $z$-direction, which sets the diagonal terms in the Hermitian permeability tensor to $\mu_{xx} = \mu_{yy} = 14\mu_0$, $\mu_{zz} = \mu_0$ and the off-diagonal terms to $\mu_{xy} = (-12.4\ii)\mu_0$, $\mu_{xz}=\mu_{yz} = 0$, where $\mu_0$ is the vacuum permeability. The band structure for both expanded and contracted unit cells, calculated using \textsc{MIT Photonic Bands} \cite{MPB}, is shown in Fig.~\ref{fig:fig3}(a). 

\begin{table}[htb]
    \centering
    \begin{tabular}{cccc}
        \toprule
        Band \# & $\raisedchi_2$ (Contracted) & $\raisedchi_1$ (Expanded) & $\Delta \raisedchi$ = $\raisedchi_2 - \raisedchi_1$ \\
        \separatorrule
        Band 1 & $(0 \, | \, 0, 0, 0)$ & $(0 \, | \, -1, -1, 0)$ & $(0 \, | \, 1, 1, 0)$\\
        Band 2 & $(0 \, | \, 0, 0, 0)$ & $(0 \, | \, 1, 1, 0)$ & $(0 \, | \, -1, -1, 0)$\\
        Band 3 & $(1 \, | \, -1, -1, -1)$ & $(1 \, | \, 0, 0, -1)$ & $(0 \, | \, -1, -1, 0)$\\
        Band 4 & $(1 \, | \, 1, 1, 1)$ & $(1 \, | \, 0, 0, 1)$ & $(0 \, | \, 1, 1, 0)$\\
        \bottomrule
    \end{tabular}
    \caption{$\chi$ indices and $\Delta \chi$ for the first four TM bands of the PhCs with the contracted and expanded unit cell types.}
    \label{tab:Chern_PhC_chi_indices}
\end{table}

Using the inversion eigenvalues at HSPs for contracted and expanded unit cell types, we determine the $\raisedchi$ indices and $\Delta \raisedchi$ for the first four TM bands (Table~\ref{tab:Chern_PhC_chi_indices}). This analysis shows that bands 1 and 2 are atomic limit bands, with Wannier centers at the $1a$ $(1b)$ position for the contracted (expanded) unit cell. Band 3 acquires a Chern number of $+1$ for both unit cell types, as can be seen from the windings in $p_x(k_y)$ shown in Fig.~\ref{fig:fig3}(b) and (c). Similarly, band 4 has a Chern number of $+1$ and is not Wannierizable. 
For all four bands, the difference in polarization between the contracted and expanded PhCs, $\Delta\mathbf{P}$, is equal to $\frac{1}{2}(\mathbf{a}_1 + \mathbf{a}_2)$, and the relative corner charge index, $\Delta Q$, is equal to $\frac{1}{2}$, as indicated by their $\Delta \raisedchi$ in Table~\ref{tab:Chern_PhC_chi_indices}.

To explore the bulk-boundary correspondence associated with the topological indices $\Delta\mathbf{P}$ and $\Delta Q$ for the Chern bands in this system, we first simulate a configuration consisting of an inner region with the expanded unit cell and an outer region consisting of the contracted unit cell, similar to the schematic in Fig.~\ref{fig:fig1}(f). In Fig.~\ref{fig:fig3}(e), we observe that the frequency spectrum contains polarization-induced non-chiral edge states, as those found in the tight-binding model in Fig.~\ref{fig:fig1}(f). We note that these edge states have been previously reported in PhC- and waveguide-based systems and may be useful for certain photonic applications \cite{chen2019strong, piccioli2022populating}. 

Next, we simulate a finite system in a ``core-cladding'' type of geometry and find corner states as shown in Fig.~\ref{fig:fig3}(f). Using a filling anomaly argument~\cite{Rot_HOTI1}, we show that both the edge and corner states originate from multiple bands and therefore have a topological origin~\cite{vaidya2023topological}. The state counting and a schematic of the photonic density of states (DOS) for the first four TM bands are shown in Fig.~\ref{fig:fig3}(g), for the structure shown in Fig.~\ref{fig:fig3}(f). This structure consists of $11\times 11$ unit cells and therefore we expect to find $121$ states per band. For bands 1 and 2, which are atomic limits with non-vanishing $\Delta \mathbf{P}$ and $\Delta Q$, we find that some of these states reside in the bandgap as edge and corner states and further that these boundary states originate from both bands 1 and 2. This leads to a mismatch between the expected number of states per band and the actual number of bulk states \cite{vaidya2023topological}. Bands 3 and 4 have a non-zero Chern number but have identical values of $\Delta \mathbf{P}$ and $\Delta Q$ to those of bands 1 and 2. As expected from these indices, we observe identical state counts of the edge and corner states that lie in the bandgap between bands 3 and 4. The observed boundary states originating from Chern bands thus clearly demonstrate the meaningfulness of $\Delta \mathbf{P}$ and $\Delta Q$ as weak topological invariants in Chern insulators.

\begin{figure}[t]
    \centering
    \includegraphics[width=\columnwidth]{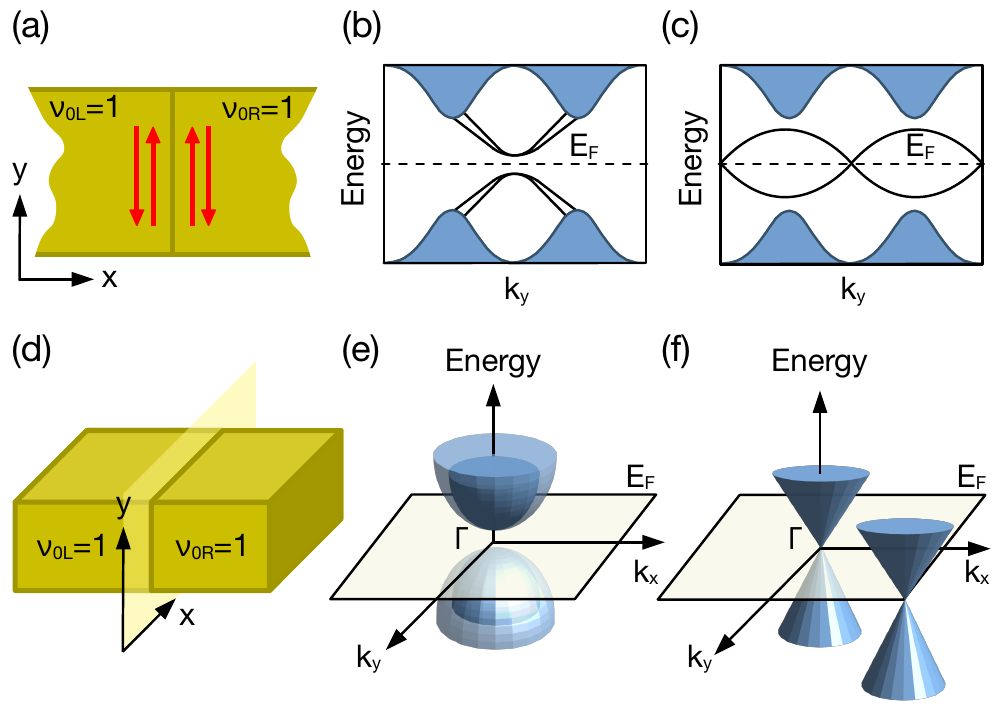}
    \caption{(a) An interface between two QSH insulators with the same non-trivial $\mathbb{Z}_2$ invariant can host either (b) a gapped trivial phase where the two pairs of helical edge states cross at the same TRIM point and hybridize or (c) a gapless non-trivial phase where the two pairs of helical edge states cross at different TRIM points.
    (d) An interface between two 3D TIs with the same strong invariant can host either (e) a gapped trivial phase where the two surface Dirac cones sit at the same TRIM point and hybridize or (f) a gapless non-trivial phase where the two surface Dirac cones sit at different TRIM points. $E_F$ marks the Fermi level.}
    \label{fig:fig4}
\end{figure}

The weak topological phenomenon discussed so far can be extended to other systems, as shown in Figure~\ref{fig:fig4}(a) where we consider the interface between two quantum spin-Hall (QSH) insulators. Each QSH insulator has a non-trivial $\mathbb{Z}_2$ invariant, $\nu_0 = 1$, and its boundary with vacuum hosts helical edge states that are Kramers-degenerate at time-reversal invariant momentum (TRIM) points \cite{bernevig2006quantum, kane2005z}. However, the interface between the two QSH insulators, at half filling, can either exhibit a trivial gapped phase, where the helical edge states of each QSH insulator cross and hybridize at the same TRIM point [Fig.~\ref{fig:fig4}(b)], or a (weak) topological gapless phase, where the helical edge states cross at different TRIM points [Fig.~\ref{fig:fig4}(c)]. Although the two QSH insulators individually have a non-trivial strong $\mathbb{Z}_2$ invariant, the presence or absence of boundary states at their common interface is characterized by the time-reversal polarization, which is a weak $\mathbb{Z}_2$ invariant \cite{fu2006time}.

Another example is obtained by considering two three-dimensional topological insulators (3D TIs), each with the same non-trivial strong index, $\nu_0$. The surface of each 3D TI (with vacuum on the exterior) is gapless and hosts a single Dirac point at one of four possible TRIM points in the surface Brillouin zone \cite{fu2007topological}. Similar to the previous example, the interface between the two strong 3D TIs, as shown in Fig.~\ref{fig:fig4}(d), hosts two Dirac points and can either be gapped or gapless at half-filling. The former case results in a trivial interface that occurs when the surface Dirac points are located at the same TRIM point and can hybridize [Fig.~\ref{fig:fig4}(e)]. The latter case is (weak) topological and occurs when the surface Dirac points are located at different TRIM points [Fig.~\ref{fig:fig4}(f)].

In this work, we have argued that the concept of electric polarization in Chern insulators is meaningful because, although mathematically ill-defined in the bulk of the material, its physical manifestations are correctly captured by the well-defined spatial and temporal \emph{difference in polarization} within Chern insulators. We have also extended the theory of inversion-symmetric weak topological phases and explored the appearance of edge and corner states in Chern insulators. Finally, we have suggested a generalization of this phenomenon to QSH insulators and 3D TIs. Our results may be experimentally probed in various platforms, such as the proposed microwave photonic crystals~\cite{wang2009observation}, optical waveguide arrays~\cite{rechtsman2013photonic, piccioli2022populating}, coupled ring resonators~\cite{hafezi2013imaging}, or electronic Chern/QSH/TI systems with grain boundaries.

We note that previous works have found other manifestations of bound charge in Chern insulators, such as at dislocations and disclinations, where it was found that weak indices play a role~\cite{disclination1, disclination2, disclination3, charge_polarization_dislocation}.

We acknowledge fruitful discussions with Thomas Christensen, Ali Ghorashi and Marius J{\"u}rgensen. M.C.R.\ and S.V.\ acknowledge the support of the U.S. Office of Naval Research (ONR) Multidisciplinary University Research Initiative (MURI) under Grant No.~N00014-20-1-2325 as well as the Charles E. Kaufman Foundation under Grant No.~KA2020-114794. W.A.B.\ acknowledges the support of startup funds from Emory University.

\bibliography{Chern_pol}

\end{document}